\documentclass[journal,onecolumn,letterpaper]{IEEEtran}
\usepackage{cite}
\usepackage{amsmath,amssymb,amsfonts}
\usepackage{graphicx}
\usepackage{textcomp}

\def\BibTeX{{\rm B\kern-.05em{\sc i\kern-.025em b}\kern-.08em
    T\kern-.1667em\lower.7ex\hbox{E}\kern-.125emX}}

\usepackage{url} 
\usepackage{booktabs} 
\usepackage{algorithm} 
\usepackage{algorithmicx}
\usepackage{algpseudocode}
\usepackage{comment}
\usepackage{makecell}
\usepackage{enumitem}

\usepackage{xcolor,colortbl}
\begin{document}

\title{LLM-Driven Feature-Level Adversarial Attacks on Android Malware Detectors}

\author{
        \IEEEauthorblockN
            {
            Tianwei Lan and
            Farid Nait-Abdesselam
	        }\\
	        \IEEEauthorblockA{Université Paris Cité, France}
        }
\maketitle

\begin{abstract}
The rapid growth in both the scale and complexity of Android malware has driven the widespread adoption of machine learning (ML) techniques for scalable and accurate malware detection. Despite their effectiveness, these models remain vulnerable to adversarial attacks that introduce carefully crafted feature-level perturbations to evade detection while preserving malicious functionality. In this paper, we present LAMLAD, a novel adversarial attack framework that exploits the generative and reasoning capabilities of large language models (LLMs) to bypass ML-based Android malware classifiers.

LAMLAD employs a dual-agent architecture composed of an LLM manipulator, which generates realistic and functionality-preserving feature perturbations, and an LLM analyzer, which guides the perturbation process toward successful evasion. To improve efficiency and contextual awareness, LAMLAD integrates retrieval-augmented generation (RAG) into the LLM pipeline. Focusing on Drebin-style feature representations, LAMLAD enables stealthy and high-confidence attacks against widely deployed Android malware detection systems.

We evaluate LAMLAD against three representative ML-based Android malware detectors and compare its performance with two state-of-the-art adversarial attack methods. Experimental results demonstrate that LAMLAD achieves an attack success rate (ASR) of up to 97\%, requiring on average only three attempts per adversarial sample, highlighting its effectiveness, efficiency, and adaptability in practical adversarial settings. Furthermore, we propose an adversarial training–based defense strategy that reduces the ASR by more than 30\% on average, significantly enhancing model robustness against LAMLAD-style attacks.

\end{abstract}


\section{Introduction}
\label{sec:intro}
Smartphones have become an integral component of modern society, functioning as essential platforms for communication, productivity, and personal data management. As of the first quarter of 2025, Android maintained its dominance in the global mobile operating system market with a share of 71.88\%~\cite{Android_Stat_2025}. However, this widespread adoption, together with Android’s open-source nature, has also made it an attractive target for malicious actors. Contemporary Android malware is capable of exfiltrating sensitive information, including geolocation data, health-related records, and financial credentials. Consequently, the development of effective Android malware detection techniques has become a critical focus for both the cybersecurity industry and the academic research community.

Over the past decade, Machine Learning (ML)-based approaches have demonstrated strong performance and flexibility in Android malware detection. These approaches typically rely on features such as Application Programming Interfaces (APIs) and permission requests to perform binary malware detection or malware family classification~\cite{zhang2018novel,GCpaper}. Despite their success, prior studies have shown that ML models are susceptible to adversarial attacks~\cite{goodfellow2014explaining,szegedy2013intriguing}. In such scenarios, attackers introduce carefully designed and imperceptible perturbations to generate adversarial examples, leading the model to misclassify malicious applications as benign. Once malware is transformed into an adversarial example, it can readily evade detection, giving rise to an ongoing arms race between the advancement of adversarial attack techniques~\cite{EvadeDroid2024,Yang23Jigsaw} and the development of robust defense mechanisms~\cite{Defensive2023ICC,chen2017securedroid} in Android malware detection systems.

In this paper, we propose LAMLAD, a novel adversarial attack framework that exploits the generative and reasoning capabilities of Large Language Models (LLMs) to evade ML-based Android malware detectors. LAMLAD adopts a dual-agent architecture consisting of an LLM manipulator and an LLM analyzer. The manipulator generates realistic feature-level perturbations while preserving the core malicious functionality, whereas the analyzer steers the modification process to achieve successful evasion. Specifically, after introducing perturbations, the LLM manipulator forwards the modified feature representation of a malicious Android Package Kit (APK) to the LLM analyzer. The analyzer, which is unaware of the applied perturbations, evaluates the features and returns both its detection outcome and an explanatory rationale when the sample is classified as \emph{Malicious}. Guided by this explanation, the LLM manipulator produces an alternative adversarial variant and resubmits the modified features to the analyzer. This iterative process continues until the analyzer labels the adversarial example as \emph{Benign}. Furthermore, LAMLAD incorporates Retrieval-Augmented Generation (RAG) into the LLM pipeline to enhance efficiency and contextual understanding. By targeting Drebin~\cite{ArpSHGR14} feature representations, LAMLAD enables stealthy and high-confidence attacks against widely adopted Android malware classifiers.

We evaluate the effectiveness of LAMLAD against three representative ML-based Android malware detection models and compare its performance with two state-of-the-art adversarial attack methods~\cite{EvadeDroid2024,HIV2020}. In addition, we investigate the impact of employing different LLMs within the LAMLAD framework. Experimental results demonstrate that LAMLAD achieves an Attack Success Rate (ASR) of 97\%, requiring an average of only 3 attempts per adversarial sample, thereby highlighting its effectiveness, efficiency, and adaptability in practical adversarial environments.

Moreover, we propose an adversarial training–based defense strategy that reduces the ASR by more than 30\% on average, effectively enhancing the robustness of malware classifiers against LAMLAD-driven attacks.

The rest of this paper is organized as follows. Section~\ref{sec:background} introduces the background on the APK structure, Drebin features, ML-based Android malware detection, adversarial attacks, and LLMs. Section~\ref{sec:LAMLAD} presents the threat model and the detailed design of the proposed LAMLAD framework. Section~\ref{sec:experiments} describes the experimental setup and reports the evaluation results with comprehensive analysis. Section~\ref{sec:defense} introduces an adversarial training–based defense mechanism and evaluates its effectiveness. Section~\ref{sec:discussion} discusses limitations and outlines potential future research directions, while Section~\ref{sec:related} surveys related work. Finally, Section~\ref{sec:conclusion} concludes the paper.

\section{Background}
\label{sec:background}

\begin{figure*}
    \centering
    \includegraphics[width=0.8\linewidth]{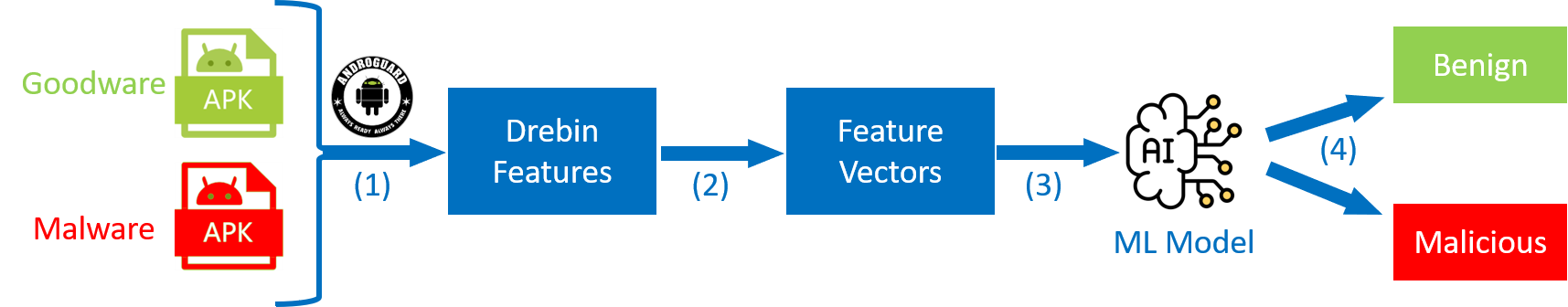}
    \caption{Training pipeline of ML-based Android malware classifiers using Drebin features}
    \label{fig:training_workflow}
\end{figure*}

\subsection{APK Structure}\label{subsec:APK}
The Android operating system distributes and installs applications in the form of APKs. An APK is a compressed archive that contains multiple files and directories required for application deployment on Android devices. These include the AndroidManifest.xml file, the classes.dex file, the lib directory (compiled native libraries), the META-INF directory (verification metadata), the assets directory (raw application resources), the res directory (uncompiled user interface resources), and the resources.arsc file (compiled resources). Since Drebin features are extracted primarily from the AndroidManifest.xml and classes.dex files, we describe these two components in detail below.

\textbf{AndroidManifest.xml} is a fundamental configuration file that specifies essential metadata about an Android application. It declares information such as the application’s name and version, required permissions, supported hardware and software features, and the set of application components. Moreover, it defines intent filters, which control how the application responds to system events or inter-application communications. Android applications are composed of four principal component types: \textit{Activities}, which manage user interfaces and handle user interactions; \textit{Services}, which execute background tasks independently of user interfaces; \textit{Content Providers}, which facilitate the sharing and management of structured data across applications; and \textit{Broadcast Receivers}, which enable applications to react to broadcast messages from the system or other applications. Permissions and intent filters play a critical role in Android’s security model by regulating access to system resources and governing interactions among components, applications, and system services.

\textbf{Classes.dex} is a compiled binary file in the Dalvik Executable (DEX) format that is executed by the Dalvik Virtual Machine or the Android Runtime. It encapsulates the application’s classes, methods, API calls, and bytecode instructions (opcodes). As the file is not directly human-readable, its analysis typically requires decompilation or reverse-engineering tools. Owing to its inclusion of the application’s core execution logic, classes.dex constitutes a primary source for static analysis techniques aimed at inferring application behavior.

\subsection{Drebin Features}\label{subsec:Drebin}
Drebin~\cite{ArpSHGR14} features constitute a widely adopted static feature representation for Android malware detection~\cite{Yang23Jigsaw,Tian23Sparsity,kim2018multimodal}. Owing to their diversity, effectiveness, and interpretability, Drebin features have been extensively used as a benchmark in both academic research and practical malware analysis. The Drebin feature set is composed of eight distinct categories: \textit{hardware components, requested permissions, application components, filtered intents, restricted API calls, used permissions, suspicious API calls, and network addresses}.

\subsection{ML-based Android Malware Detection}\label{subsec:MLAndroid}
ML-based Android malware detection has emerged as an effective and scalable solution for identifying malicious applications within the rapidly expanding Android ecosystem~\cite{zhang2018novel,kim2018multimodal,li2018significant}. In contrast to traditional signature-based detection techniques, ML-based approaches are capable of learning discriminative patterns from large collections of APKs, enabling the detection of previously unseen malware samples.

These models are commonly trained on features extracted via static, dynamic, or hybrid analysis techniques, including permissions, API calls, control flow, network behavior, or system calls. Static analysis methods extract features without executing the application, thereby offering advantages in terms of efficiency and safety. Dynamic analysis monitors application behavior during runtime and is able to capture more evasive or obfuscated malware behaviors. ML-based detection systems can support both binary classification tasks and multiclass classification tasks, often achieving high detection accuracy. A variety of feature representation strategies have been explored in ML-based Android malware detection, including text-based, graph-based, and image-based methods.

Fig. \ref{fig:training_workflow} illustrates the training workflow of ML-based Android malware classifiers using Drebin features, which is adopted in this paper. In Step~(1), Drebin features are extracted from each APK in the training dataset using Androguard~\cite{Androguard}, a widely used Android reverse-engineering tool. In Step~(2), the extracted features are encoded into high-dimensional binary feature vectors. In Step~(3), these feature vectors, together with their corresponding labels, are used to train ML classifiers. Finally, in Step~(4), the trained model is capable of accurately distinguishing between benign applications and malware.

\subsection{Adversarial Attacks}\label{subsec:AdvAttacks}     
We consider an ML model $M$ that accurately classifies a given input example $x$ as its true label $Y$, such that $M(x) = Y$. 
However, an attacker may attempt to manipulate the input by introducing a small, carefully designed perturbation $\rho$, resulting in a modified input $x^{\prime} = x + \rho$. The goal of the attacker is to deceive the model into misclassifying the adversarial example $x^{\prime}$ with high confidence~\cite{szegedy2013intriguing}. In this scenario, the perturbed input is intentionally crafted so that the model predicts a class different from $Y$, as given in Equation~\ref{eq:perturb Model}.
\begin{equation}\label{eq:perturb Model}
M(x^{\prime}) = Y^{\prime} \hspace{0.3cm} | \hspace{0.3cm} Y^{\prime}\neq Y 
\end{equation}

\textbf{Attacker’s Knowledge~\cite{Wild_Biggio}.} \textit{White-box attacks} assume full knowledge of the target model, including the training data, feature representation, model architecture, learning algorithm, and model parameters. \textit{Gray-box attacks} are conducted under partial knowledge of the target system. In contrast, \textit{black-box attacks} operate without any prior information about the target model.


        

\textbf{Attacker’s Capability~\cite{Wild_Biggio}.} \textit{Poisoning attacks} tamper with the training dataset to undermine the learning process, thereby inducing misclassification by the target model during inference. In some cases, such attacks may also involve manipulating test samples in conjunction with the poisoned training data. In contrast, \textit{evasion attacks} are restricted to modifying inputs to circumvent the detector in the model deployment phase, under the assumption that the training procedure and model parameters remain unchanged.

\begin{figure*}
    \centering
    \includegraphics[width=0.9\linewidth]{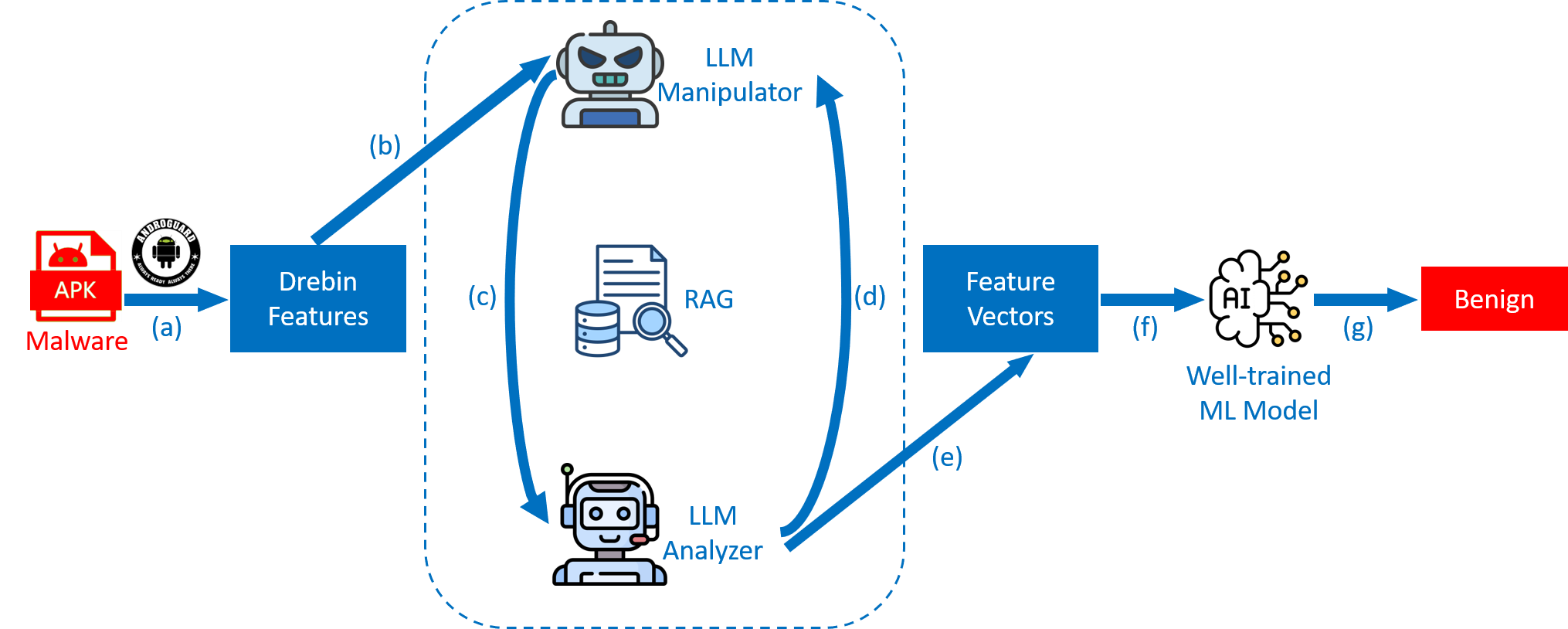}
    \caption{LAMLAD attack workflow against ML-based Android malware classifiers using Drebin features}
    \label{fig:attack_workflow}
\end{figure*}

\subsection{LLM}\label{subsec:LLM}
LLMs represent a significant advancement in the field of Natural Language Processing (NLP), enabling machines to comprehend and generate human language with a high degree of fluency and accuracy. These models are built upon the transformer architecture~\cite{Attention2017}, which employs self-attention mechanisms to more effectively capture long-range contextual dependencies within textual sequences compared to earlier ML architectures.

LLMs are typically pre-trained on massive and heterogeneous text corpora, allowing them to generalize to a wide range of downstream tasks with minimal or no task-specific fine-tuning~\cite{SurveyLLM2023}. Relative to traditional ML models, LLMs exhibit superior capabilities in both language generation and reasoning tasks, as well as improved scalability across diverse application domains.

The range of applications for LLMs continues to expand rapidly, encompassing areas such as conversational agents, machine translation, document summarization, code generation, and information retrieval. While recent studies have investigated the use of LLMs for Android malware detection~\cite{AppPoet2025,LAMD2025}, research on exploiting LLMs to construct effective adversarial attacks against Android malware classifiers remains limited.

\section{LAMLAD}\label{sec:LAMLAD}
This section presents the threat model and outlines the attack strategy of LAMLAD, detailing its architecture and operational workflow.

\subsection{Threat Model}\label{subsec:Threat_Model}
\textbf{Goal.} As discussed in Section~\ref{subsec:AdvAttacks}, the goal of LAMLAD is to mislead the target classifier into confidently assigning incorrect labels to adversarial samples. Furthermore, each malicious sample $x$ requires a distinct perturbation $\rho_{x}$ in order to successfully evade detection. In this work, we encode the classification outcomes \emph{Benign} and \emph{Malicious} as 0 and 1, respectively. Under these definitions, the attack objective can be formally expressed in Equation~\ref{eq:Goal}.
\begin{align}\label{eq:Goal}
\rho_{x}^{*} &= \operatorname*{arg\,min}_{\rho_{x}\in F} \, cost(\rho_{x}) \nonumber \\
\text{such that} \quad & M(x+\rho_{x}^{*}) = 0
\end{align}
$F$ denotes the complete set of Drebin feature perturbations, and $cost()$ represents the cost incurred by an attacker when applying a given perturbation. According to Equation~\ref{eq:Goal}, LAMLAD aims to minimize the perturbation cost for each malware sample in order to maintain stealthiness. Moreover, the modification process is constrained to the addition of Drebin features, without removing any original features, thereby preserving the malware’s inherent functionality and malicious behavior.

\textbf{Knowledge.} LAMLAD has access to the feature space and the training workflow of the target classifier. However, it does not possess knowledge of the training dataset, model architecture, or model parameters. Consequently, LAMLAD operates under a gray-box threat model.

\textbf{Capability.} Since LAMLAD modifies only the Drebin features of malicious test samples to evade detection without interfering with the training process, it is categorized as an evasion attack.



\subsection{Attack Strategy}\label{subsec:Attack_Strategy}
Building upon the standard training workflow illustrated in Fig.~\ref{fig:training_workflow}, we design LAMLAD, a novel adversarial attack framework that leverages LLMs to evade ML-based Android malware detectors. Fig.~\ref{fig:attack_workflow} presents an overview of the LAMLAD attack workflow targeting ML-based Android malware classifiers using Drebin features.

\textbf{Step (a):} LAMLAD begins by extracting Drebin features from a malicious APK in the test dataset using Androguard.

\textbf{Step (b):} The extracted Drebin features are then provided as input to an LLM manipulator, whose objective is to generate realistic feature-level perturbations. As discussed in Section~\ref{subsec:Threat_Model}, LAMLAD seeks to minimize perturbation cost while preserving the consistency of malicious behavior. To satisfy these constraints, the LLM manipulator is restricted to adding only one Drebin feature to the current feature set.

\textbf{Step (c):} The LLM manipulator forwards the modified Drebin feature set, including all original features, to an LLM analyzer. The analyzer, which is unaware of any introduced perturbations, evaluates the complete feature set and produces both a malware detection decision and an accompanying explanation.

\textbf{Step (d):} If the LLM analyzer classifies the sample as \emph{Malicious}, it returns the detection result and explanatory feedback to the LLM manipulator. Guided by this explanation, the manipulator may either append an additional Drebin feature to the existing feature set or revoke the most recent addition and substitute it with an alternative feature. Notably, this process never removes any original Drebin features of the malware.

\textbf{Step (e):} Once the LLM analyzer labels the modified sample as \emph{Benign}, the resulting feature set is encoded into a high-dimensional binary feature vector.

\textbf{Step (f):} This feature vector is subsequently provided as input to a pre-trained ML-based malware classifier.

\textbf{Step (g):} Ultimately, the target ML model misclassifies the adversarial example as \emph{Benign}, thereby completing the attack.

It is apparent that the interaction loop between the LLM manipulator and the LLM analyzer may continue indefinitely if the analyzer never classifies the adversarial example as \emph{Benign}. Therefore, we define an upper bound on the number of attempts permitted for the LLM manipulator in Section~\ref{sec:experiments}.

\begin{figure}
    \centering
    \includegraphics[width=0.5\linewidth]{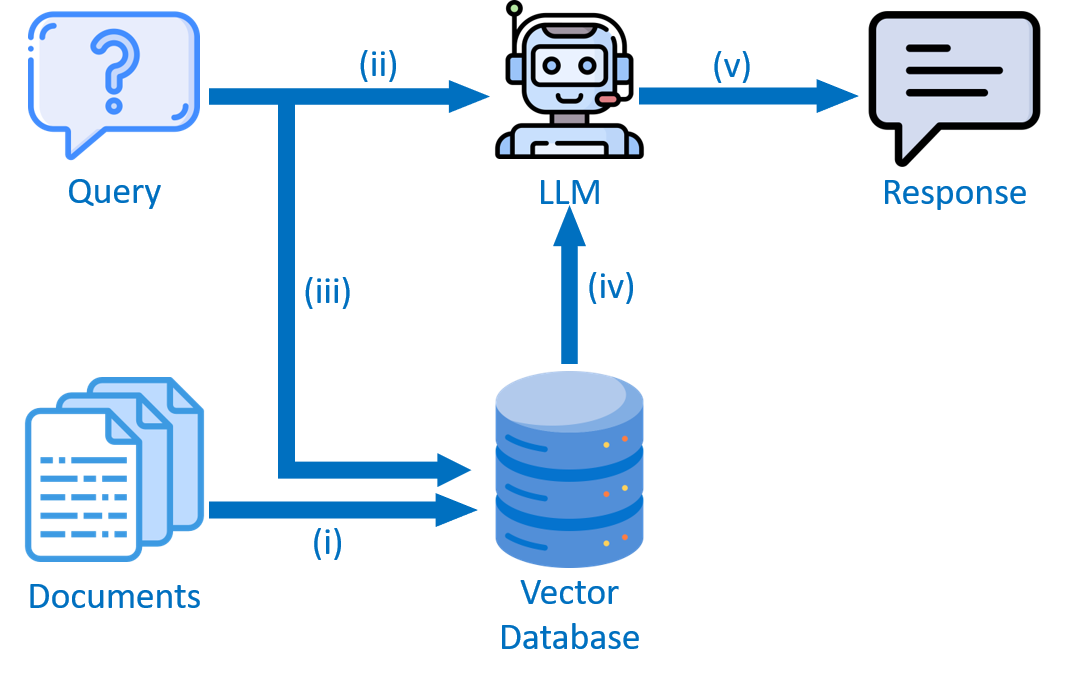}
    \caption{Overview of the RAG-enhanced interaction workflow between the LLM manipulator and LLM analyzer in LAMLAD}
    \label{fig:RAG_workflow}
\end{figure}

To enhance efficiency and contextual awareness, LAMLAD incorporates the Retrieval-Augmented Generation (RAG)~\cite{RAG2020} mechanism into Steps~(b), (c), and (d) of Fig.~\ref{fig:attack_workflow}. RAG augments LLM inference by retrieving task-relevant information from an external knowledge base and integrating it into the input prompt at inference time. This approach enables LLMs to produce more accurate and contextually grounded outputs without requiring model retraining. Within LAMLAD, the objective of integrating RAG is twofold: to strengthen the LLM manipulator’s ability to generate effective feature perturbations and to improve the accuracy and explanatory quality of malware detection performed by the LLM analyzer. The RAG-enhanced interaction workflow between the LLM manipulator and the LLM analyzer is illustrated in Fig.~\ref{fig:RAG_workflow}.

\textbf{Step (i):} To construct the external knowledge base, we first collect comprehensive documentation of Drebin features from the Android Developers website~\cite{AndroidDev}, including feature descriptions, usage guidelines, and illustrative examples. These documents are segmented into semantically coherent chunks at the paragraph level. Each chunk is subsequently transformed into a dense vector representation using \textit{all-MiniLM-L6-v2}~\cite{Embedding}, a pre-trained sentence embedding model. The resulting vectors are indexed and stored in a \textit{FAISS}~\cite{FAISS} vector database. \textit{FAISS} (Facebook AI Similarity Search) is an open-source library designed for efficient similarity search and clustering over large-scale dense vector collections.

\textbf{Step (ii):} During inference, a natural language query is constructed as part of the final prompt. In the LAMLAD framework, this query comprises the Drebin features used in Steps~(b)-(d), as well as the additional explanatory information required in Step~(d) of Fig.~\ref{fig:attack_workflow}.

\textbf{Step (iii):} The query also serves as the input for the retrieval process. To improve retrieval efficiency and reduce the input token consumption of the LLM, the query is partitioned into multiple groups, with each group containing Drebin features belonging to the same category or class. For example, the API call \textit{"android.telephony.SmsManager.sendTextMessage()"} is grouped under the \textit{"android.telephony.SmsManager"} class. Each group is converted into a vector embedding using \textit{all-MiniLM-L6-v2}, which is then used to perform a nearest-neighbor search over the vector database. For each group, the top-5 most similar vectors are retrieved based on their L2 distances. The associated text chunks are treated as the most relevant contextual information for that group.

\textbf{Step (iv):} All retrieved chunks are concatenated to form a unified context window that augments the original query. This enriched context is combined with the query to construct the final prompt.

\textbf{Step (v):} The finalized prompt is then provided to the LLM, which generates the corresponding response.

Table~\ref{tab:prompt_templates} presents the prompt templates used in Steps~(b)-(d) of Fig.~\ref{fig:attack_workflow}. Each prompt is composed of three main components: the system setup, the contextual information (\textit{\{CONTEXT\}}), and the user query (\textit{\{QUERY\}}). In addition, the prompt employed in Step~(d) includes the most recently added feature generated by the LLM manipulator (\textit{\{LAST ADDED FEATURE\}}).
\begin{table}
    \definecolor{lightgray}{gray}{0.92}
    \caption{Prompt templates used in Steps~(b)-(d) of Fig.~\ref{fig:attack_workflow}}
    \label{tab:prompt_templates}
    \centering
\resizebox{0.6\linewidth}{!}
{
    \begin{tabular}{ccc}
\specialrule{1.2pt}{0pt}{0pt}
      \textbf{Step}   & \textbf{LLM} & \textbf{Prompt} \\
\hline
\rowcolor{lightgray}
       (b) & manipulator & \makecell[l]{You are an Android malware manipulator.\\ Your goal is to evade ML-based Android \\ malware detection. Please add only one \\ Drebin feature to the input features based \\ on the context below. The output should \\ be the modified feature set including \\ input features. \\ context: \textit{\{CONTEXT\}} \\ input features: \textit{\{QUERY\}}} \\
       (c) & analyzer & \makecell[l]{You are an Android malware analyzer.\\ Your goal is to detect malware based on \\ the input Drebin features and the context \\ below. If it is malware, the output should \\ be "Malicious" with the explanation. \\ If it is goodware, the output should be \\ "Benign" with the explanation. \\ context: \textit{\{CONTEXT\}} \\ input features: \textit{\{QUERY\}}} \\
\rowcolor{lightgray}
       (d) & manipulator & \makecell[l]{You are an Android malware manipulator.\\ Your goal is to evade ML-based Android \\ malware detection. An Android expert \\ classifies the APK as malware with its \\ explanation based on the input features. \\ Informed by the input features with \\ explanation and the context below, you can: \\ (1) either add only one new Drebin feature \\ to the input features, (2) or remove the last \\ added feature \textit{\{LAST ADDED FEATURE\}} \\ and add only one different Drebin feature. \\ The output should be the modified feature \\ set including input features. \\ context: \textit{\{CONTEXT\}} \\ input features and explanation: \textit{\{QUERY\}}} \\
\specialrule{1.2pt}{0pt}{0pt}
    \end{tabular}
}
\end{table}

Algorithm~\ref{algo:LAMLAD} illustrates the adversarial example generation procedure of the LAMLAD framework. The initialization of all variables is performed in Line~1, while Line~2 corresponds to Step~(a) in Fig.~\ref{fig:attack_workflow}. The iterative interaction between the LLM manipulator and the LLM analyzer is described in Lines~3-19. Specifically, the LLM manipulator generates feature perturbations in Lines~4-11, and the LLM analyzer produces the detection decision along with the corresponding explanation in Lines~12-13. Lines~14-18 determine whether the interaction loop should continue. Upon termination of the loop, the resulting adversarial feature set is embedded into a feature vector, which is subsequently provided as input to an ML-based classifier for final prediction.

Algorithm~\ref{algo:LAMLAD} employs four core functions. \textsc{ANDROGUARD}() extracts the original Drebin feature set from a malicious APK using Androguard. \textsc{RAG}() retrieves relevant contextual information for a given query in accordance with Steps~(ii)-(iv) shown in Fig.~\ref{fig:RAG_workflow}. \textsc{MANIPULATOR}() generates the modified Drebin feature set based on the inputs specified in Steps~(b) or (d) of Table~\ref{tab:prompt_templates}. Finally, \textsc{ANALYZER}() outputs the detection result and the corresponding explanation using the retrieved context and the modified feature set, as illustrated in Step~(c) of Table~\ref{tab:prompt_templates}.


\begin{algorithm}[!htb]
\caption{LAMLAD}\label{algo:LAMLAD}
\begin{algorithmic}[1]
\Statex \textbf{Constants:} the malware $apk$, the maximum number of attempts $A$
\Statex \textbf{Variables:} the original Drebin feature set of a malicious APK $x$, the perturbation feature set $\rho$, the last added feature $f$, the output feature set of LLM manipulator $x^{\prime}$, the context $cxt$, the number of attempts $a$, the detection result $r$ and explanation $e$ of LLM analyzer
\Statex
\State \textbf{initialize} $\rho \leftarrow \{\}$, $a \leftarrow 1$, $f \leftarrow \emptyset$, $e \leftarrow \emptyset$
\State $x \leftarrow$ ANDROGUARD$(apk)$
\While{$a \le A$}
    \State $cxt \leftarrow$ RAG($x+\rho+e$)
    \State $x^{\prime} \leftarrow$ MANIPULATOR($f$, $cxt$, $x+\rho+e$)
    \If{$f$ \textbf{in} $x^{\prime}$}
        \State $f \leftarrow x^{\prime}-x-\rho$
    \Else
        \State $f \leftarrow x^{\prime}-x-(\rho-f)$
    \EndIf
    \State $\rho \leftarrow x^{\prime}-x$
    \State $cxt \leftarrow$ RAG($x^{\prime}$)
    \State $r,e \leftarrow$ ANALYZER($cxt$, $x^{\prime}$)
    \If{$r=$"Benign"}
        \State \textbf{break}
    \Else
        \State $a \leftarrow a+1$
    \EndIf
\EndWhile
\State \textbf{return} $x^{\prime}$
\end{algorithmic}
\end{algorithm}

\section{Experiments and Evaluations}\label{sec:experiments}
In this section, we conduct a comprehensive experimental evaluation to assess the effectiveness of LAMLAD against ML-based Android malware detection systems. First, the experimental setup is described in Section~\ref{subsec:Experiment Setup}. Next, Section~\ref{subsec:Performance ML Detectors} reports the baseline performance of ML-based malware detectors in the absence of adversarial attacks. Subsequently, Section~\ref{subsec:Performance LLM with RAG} examines the performance of LLMs augmented with RAG for malware detection, demonstrating the effectiveness of both the RAG mechanism and the LLM analyzer in comparison with traditional ML detectors. Finally, Section~\ref{subsec:Performance LAMLAD} presents the results of LAMLAD attacks against ML-based detectors and compares its performance with two state-of-the-art adversarial attack methods.

\subsection{Experiment Setup}\label{subsec:Experiment Setup}
\textbf{Dataset.}
We adopt the dataset constructed by Ruggia et al.~\cite{ruggia24unmasking}, which comprises 20K malicious APKs and 20K benign APKs. The malicious samples span 196 malware families, with each family containing approximately 100 to 110 APKs. The benign samples cover 50 application categories obtained from the Google Play Store~\cite{GooglePlayStore}. To mitigate the impact of class imbalance on classification performance, we maintain an equal number of benign and malicious samples. In addition, dataset diversity is preserved to promote the generalization capability of ML models and enhance their robustness across different malware families.
The complete dataset is partitioned into 80\% for training and 20\% for test in the evaluation of ML-based malware classifiers presented in Section~\ref{subsec:Performance ML Detectors}. For consistency and fair comparison, the same test set is subsequently used for evaluating LLM-based malware detection and the LAMLAD attack framework in Sections~\ref{subsec:Performance LLM with RAG} and~\ref{subsec:Performance LAMLAD}, respectively.

\textbf{ML Models.}
We consider three representative ML-based models for Android malware detection~\cite{ArpSHGR14,dambra2023decoding}, all trained using Drebin feature representations. The first model is a Support Vector Machine (SVM) with a linear kernel and a regularization parameter of $\lambda = 1$. The second model is a Gradient Boosted Trees (GBT) classifier composed of 3000 trees. The third model is a feed-forward Neural Network (NN) with five layers containing 24{,}000, 240, 120, 60, and 1 neurons, respectively. Rectified Linear Unit (ReLU) activation functions are employed in the hidden layers, while a Sigmoid activation function is used in the output layer. The NN produces a single probability value indicating the likelihood that an input sample is malicious.
Given the extremely high dimensionality of Drebin feature vectors (732{,}800 dimensions), we apply Random Projections~\cite{Li2006kdd} to reduce the input dimensionality to 24{,}000 prior to NN training. The NN is optimized using the Adam optimizer~\cite{kingma2017adam}, with cross-entropy loss adopted as the training objective.

\textbf{LLMs.}
In this study, we select three state-of-the-art and widely adopted LLMs: GPT-4o~\cite{GPT-4o}, Gemini-2.0-Flash~\cite{Gemini-2.0}, and DeepSeek-V3~\cite{DeepSeek-V3}. GPT-4o (version: 2024-11-20), developed by OpenAI, belongs to the GPT-4 model family and supports both textual and visual inputs. It comprises over 175 billion parameters and provides a maximum context window of 128K tokens, with a rate limit of 30K tokens per minute for tier-1 users. Gemini-2.0-Flash (version: 2025-02-05), developed by Google DeepMind, supports multimodal inputs including text, images, audio, and video. It offers an extended context window of up to 1M tokens and a rate limit of 4M tokens per minute for tier-1 users. DeepSeek-V3 (version: 2024-12-26), developed by DeepSeek, is a mixture-of-experts LLM with 671 billion total parameters, of which 37 billion are activated per token during inference. DeepSeek-V3 supports a context window of 128K tokens and imposes no explicit rate limit.
Based on the prompt templates presented in Table~\ref{tab:prompt_templates}, all prompts used in our experiments fall well within the maximum context window supported by each selected LLM. Furthermore, we set the temperature parameter to 0 for all models to promote deterministic and reproducible outputs.

\textbf{Metrics.} 
To evaluate the performance of ML-based detectors and LLM-based approaches for malware detection in Sections~\ref{subsec:Performance ML Detectors} and~\ref{subsec:Performance LLM with RAG}, we report the following metrics: (1) the True Positive Rate (TPR), which measures the proportion of malware samples correctly identified; (2) the False Positive Rate (FPR), defined as the proportion of benign applications incorrectly classified as malware; (3) the overall classification Accuracy (ACC); and (4) the F1-score (F1), representing the harmonic mean of precision and recall.
In Section~\ref{subsec:Performance LAMLAD}, we adopt the Attack Success Rate (ASR) to quantify the proportion of adversarial examples that successfully evade ML-based detection among all malicious samples subjected to attack. In addition, we report the number of attempts required by the LLM manipulator to generate an adversarial example.

\subsection{Performance of ML Detectors}\label{subsec:Performance ML Detectors}
Following the training workflow illustrated in Fig.~\ref{fig:training_workflow}, we first evaluate the performance of ML-based malware detectors in the absence of adversarial attacks, establishing baseline results for comparison with LLM-based malware detection in Section~\ref{subsec:Performance LLM with RAG}. These baseline results also demonstrate that the ML detectors are properly trained and effective for malware classification prior to the application of adversarial attacks, as discussed in Section~\ref{subsec:Performance LAMLAD}. Table~\ref{tab:ML detectors} reports the detection performance of all three ML models, namely SVM, GBT, and NN.

\begin{table}
    \definecolor{lightgray}{gray}{0.92}
    \caption{Detection performance of three ML-based malware detectors}
    \label{tab:ML detectors}
    \centering
\resizebox{0.6\linewidth}{!}
{
    \begin{tabular}{ccccc}
\specialrule{1.2pt}{0pt}{0pt}
        \textbf{Model} & \textbf{TPR(\%)} & \textbf{FPR(\%)} & \textbf{ACC(\%)} & \textbf{F1(\%)}\\
\hline
\rowcolor{lightgray}
        SVM & 97.24 & 3.26 & 97.00 & 96.96\\
        GBT & 97.58 & 2.07 & 97.75 & 97.72\\
\rowcolor{lightgray}
        NN & 96.94 & 3.40 & 96.73 & 96.69\\
\specialrule{1.2pt}{0pt}{0pt}
    \end{tabular}
}
\end{table}

The results show that all three ML models exhibit strong detection performance on the test dataset. Each model achieves a TPR exceeding 96.9\%, with the GBT model attaining the highest TPR of 97.58\%. In addition, all models maintain relatively low FPRs, ranging from 2.07\% to 3.4\%. The overall accuracy of each model surpasses 96.7\%, while consistently high F1-scores indicate a favorable balance between precision and recall. Collectively, these results confirm that all three ML detectors are well trained and effective at discriminating between benign and malicious APKs in the test set.

\subsection{Performance on Malware Detection Using LLMs with RAG}\label{subsec:Performance LLM with RAG}
To evaluate the effectiveness of both the RAG mechanism and the LLM analyzer, we investigate the performance of LLMs augmented with RAG for Android malware detection. As described in Step~(c) of Fig.~\ref{fig:attack_workflow} and in Table~\ref{tab:prompt_templates}, we employ three LLMs, GPT-4o, Gemini-2.0-Flash, and DeepSeek-V3, to generate malware detection decisions (\emph{Malicious} or \emph{Benign}) along with corresponding explanations, using Drebin feature representations of APKs as input. We use the same test set comprising 8K samples and adopt the same evaluation metrics as those used in Section~\ref{subsec:Performance ML Detectors} to ensure a fair comparison.
To assess the impact of RAG, each LLM is evaluated both with and without RAG integration. Table~\ref{tab:LLM detectors} summarizes the malware detection performance of the three LLMs under these two settings.

\begin{table}
    \definecolor{lightgray}{gray}{0.92}
    \caption{Detection performance of three LLMs with and without RAG}
    \label{tab:LLM detectors}
    \centering
\resizebox{0.7\linewidth}{!}
{
    \begin{tabular}{ccccc}
\specialrule{1.2pt}{0pt}{0pt}
        \textbf{LLM} & \textbf{TPR(\%)} & \textbf{FPR(\%)} & \textbf{ACC(\%)} & \textbf{F1(\%)}\\
\hline
\rowcolor{lightgray}
        GPT & 79.80 & 16.78 & 81.51 & 81.19\\
        GPT+RAG & 97.35 & 2.43 & 97.46 & 97.46\\
\rowcolor{lightgray}
        Gemini & 81.43 & 14.10 & 83.66 & 83.29\\
        Gemini+RAG & 97.83 & 1.78 & 98.03 & 98.02\\
\rowcolor{lightgray}
        DeepSeek & 75.98 & 19.93 & 78.03 & 77.57\\
        DeepSeek+RAG & 92.43 & 4.38 & 94.03 & 93.93\\
\specialrule{1.2pt}{0pt}{0pt}
    \end{tabular}
}
\end{table}

When used without external augmentation, all three LLMs exhibit relatively limited malware detection performance. Among them, Gemini slightly outperforms GPT, achieving a TPR of 81.43\% and an FPR of 14.10\%. In contrast, DeepSeek demonstrates the weakest performance, with the lowest TPR of 75.98\% and the highest FPR of 19.93\%, indicating reduced detection sensitivity and a higher propensity for false positives.

The integration of RAG leads to substantial improvements across all evaluated metrics for each LLM. Specifically, GPT+RAG increases its TPR from 79.80\% to 97.35\% while reducing its FPR from 16.78\% to 2.43\%, accompanied by notable gains in accuracy and F1-score. Similar performance enhancements are observed for Gemini and DeepSeek. In particular, Gemini+RAG achieves the strongest overall results, with a TPR of 97.83\%, an FPR of 1.78\%, and an accuracy of 98.03\%. DeepSeek+RAG also benefits markedly from RAG integration, reducing its FPR by more than 15\% and increasing its TPR by over 16\%. With RAG enabled, all LLMs attain F1-scores exceeding 93.9\%.

A comparison between Table~\ref{tab:LLM detectors} and Table~\ref{tab:ML detectors} further indicates that RAG-equipped LLMs achieve detection performance comparable to that of traditional ML-based detectors. Notably, Gemini+RAG outperforms all three evaluated ML models, while GPT+RAG (F1-score: 97.46\%) surpasses both the SVM and NN classifiers.

Overall, these results demonstrate that RAG substantially enhances the effectiveness of LLMs by incorporating relevant external knowledge during inference. This integration not only mitigates hallucination effects but also improves factual consistency and malware detection accuracy. Moreover, the explanations generated by the LLMs remain consistent with their detection decisions, confirming the effectiveness and reliability of the LLM analyzer in Section~\ref{subsec:Performance LAMLAD}.

\subsection{Performance of LAMLAD}\label{subsec:Performance LAMLAD}
To assess the effectiveness of LAMLAD, we present and analyze the results of LAMLAD attacks against ML-based Android malware detectors. Based on the findings in Section~\ref{subsec:Performance LLM with RAG}, we select the two best-performing LLMs integrated with RAG, namely Gemini+RAG and GPT+RAG, for subsequent experiments. For simplicity, as all LLMs are coupled with RAG in the following evaluations, we refer to each LLM+RAG configuration simply as an LLM.
Following the architecture illustrated in Fig.~\ref{fig:attack_workflow}, the selected LLMs are assigned distinct roles within the LAMLAD framework. This setup yields four possible \emph{Manipulator-Analyzer} configurations: GPT-GPT, GPT-Gemini, Gemini-GPT, and Gemini-Gemini. We emphasize that the LLM manipulator and the LLM analyzer operate as two independent LLM instances, even when the same LLM type is used in configurations such as GPT-GPT and Gemini-Gemini.
The LAMLAD attack is evaluated against three ML-based malware detectors, namely SVM, GBT, and NN. To ensure a reliable measurement of the ASR, only malicious samples that are initially classified as \emph{Malicious} (i.e., true positives) by the target ML detectors are selected for adversarial example generation. This selection strategy avoids the inclusion of false negatives, which could otherwise bias the ASR evaluation.

\begin{figure}
    \centering
    \includegraphics[width=0.7\linewidth]{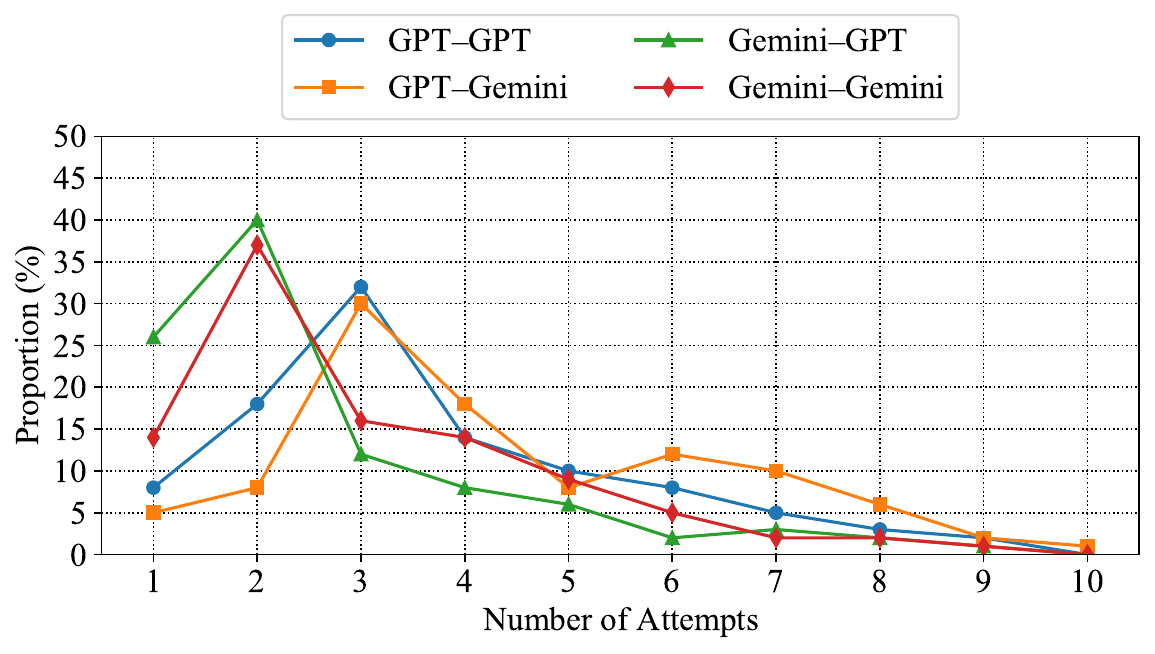}
    \caption{Distribution of the number of attempts}
    \label{fig:NB_Attempts}
\end{figure}

We first analyze the number of attempts required by the LLM manipulator to generate adversarial examples. Fig.~\ref{fig:NB_Attempts} presents the distribution of attack attempts, where the x-axis represents the number of attempts and the y-axis denotes the proportion of adversarial examples generated at each attempt count. The results indicate that the majority of adversarial examples are produced within the initial few attempts. Notably, the Gemini-GPT configuration achieves the highest proportion at the second attempt (40\%), followed closely by Gemini-Gemini (37\%). In contrast, GPT-GPT and GPT-Gemini reach their respective peaks at the third attempt, with proportions of 32\% and 30\%.
Beyond these peak points, the proportion of adversarial examples decreases steadily as the number of attempts increases, with the exception of a minor increase observed for GPT-Gemini at the sixth attempt. For most configurations, fewer than 3\% of adversarial examples are generated after the eighth attempt, and the proportions approach zero by the ninth and tenth attempts. These observations suggest that extended manipulation cycles are rarely necessary, as successful perturbations typically emerge early in the attack process. Additionally, fewer than 0.8\% of malicious samples fail to be converted into adversarial examples even after 20 attempts. To emphasize the representative distributions, these rare cases are excluded from Fig.~\ref{fig:NB_Attempts}.
We further compute the average number of attempts required for each \emph{Manipulator-Analyzer} pairing: 3.71 for GPT-GPT, 4.41 for GPT-Gemini, 2.62 for Gemini-GPT, and 3.06 for Gemini-Gemini. Among these, the Gemini-GPT pairing requires the fewest attempts on average, indicating the highest efficiency in rapidly generating adversarial examples. Overall, these findings highlight the efficiency of the \emph{Manipulator-Analyzer} configurations, as reflected by both the early peak distributions and an average of approximately three attempts per adversarial example.

We next evaluate the ASR of LAMLAD against three ML-based malware detectors, namely SVM, GBT, and NN. Based on the observations from Fig.~\ref{fig:NB_Attempts}, we limit the LLM manipulator to a maximum of 10 attempts per adversarial example. We then compare the performance of the four LAMLAD configurations with two state-of-the-art adversarial attack techniques, HIV~\cite{HIV2020} and EvadeDroid~\cite{EvadeDroid2024}, which are representative evasion attacks against Android malware detection and are introduced in Section~\ref{sec:related}. All evaluated attacks operate under the same attacker's knowledge assumptions. Table~\ref{tab:ASR} summarizes the ASR achieved by the different adversarial attack methods against the three ML detectors.

\begin{table}
    \definecolor{lightgray}{gray}{0.92}
    \caption{ASR (\%) of different adversarial attack methods against three ML detectors}
    \label{tab:ASR}
    \centering
\resizebox{0.5\linewidth}{!}
{
    \begin{tabular}{cccc}
\specialrule{1.2pt}{0pt}{0pt}
        \textbf{Method} & \textbf{SVM} & \textbf{GBT} & \textbf{NN}\\
\hline
\rowcolor{lightgray}
        EvadeDroid & 81.94 & 79.77 & 85.18 \\
        HIV & 91.42 & 90.65 & 93.33 \\
\hline
\rowcolor{lightgray}
        GPT–GPT & 88.76 & 86.21 & 91.29 \\
        GPT–Gemini & 94.37 & 93.72 & 94.90 \\
\rowcolor{lightgray}
        Gemini–GPT & 95.86 & 95.09 & 96.71 \\
        Gemini–Gemini & \textbf{97.10} & \textbf{96.68} & \textbf{97.52} \\
\specialrule{1.2pt}{0pt}{0pt}
    \end{tabular}
}
\end{table}

Overall, all LAMLAD configurations substantially outperform EvadeDroid, achieving ASR improvements ranging from approximately 6\% to more than 16\% across all evaluated ML detectors. When compared with HIV, LAMLAD exhibits comparable or superior performance, particularly for configurations involving Gemini-based pairings. Among all LAMLAD variants, the Gemini-Gemini configuration attains the highest ASR across the three ML detectors (SVM: 97.10\%, GBT: 96.68\%, NN: 97.52\%), followed closely by the Gemini-GPT and GPT-Gemini pairings. Although the GPT-GPT configuration yields the lowest ASR among the LAMLAD variants, it still consistently outperforms EvadeDroid and achieves performance close to that of HIV.

Despite the strong baseline effectiveness of ML-based malware detectors, LAMLAD is able to achieve an ASR of up to 97\% using the Gemini-Gemini pairing. Combined with an average of approximately three attempts required to generate each adversarial example, these results underscore the effectiveness, efficiency, and adaptability of LAMLAD in practical adversarial scenarios targeting ML-based Android malware detection systems.

\section{Defense Against LAMLAD}\label{sec:defense}
Based on the results of the preceding experiments, LAMLAD highlights the effectiveness of LLM-driven adversarial strategies, underscoring the urgent need to develop robust defense mechanisms for ML-based malware detectors against such evolving threats. Adversarial training~\cite{goodfellow2014explaining} is a widely adopted defense approach that enhances model robustness by incorporating adversarial examples into the training process. Specifically, adversarial training augments the original training dataset with adversarial samples labeled with their correct classes, thereby encouraging the model to learn more robust decision boundaries and to better differentiate between legitimate and manipulated inputs. Consequently, the trained model becomes more resilient to adversarial perturbations and less prone to misclassification when exposed to adversarial examples during inference.
In our defense evaluation, we integrate adversarial training by augmenting the original training dataset with 50 randomly selected adversarial examples from each LAMLAD \emph{Manipulator-Analyzer} pairing. These 200 adversarial examples are all labeled as \emph{Malicious} and are jointly used with the original training data to retrain the ML-based malware detectors.

\begin{figure}
    \centering
    \includegraphics[width=0.7\linewidth]{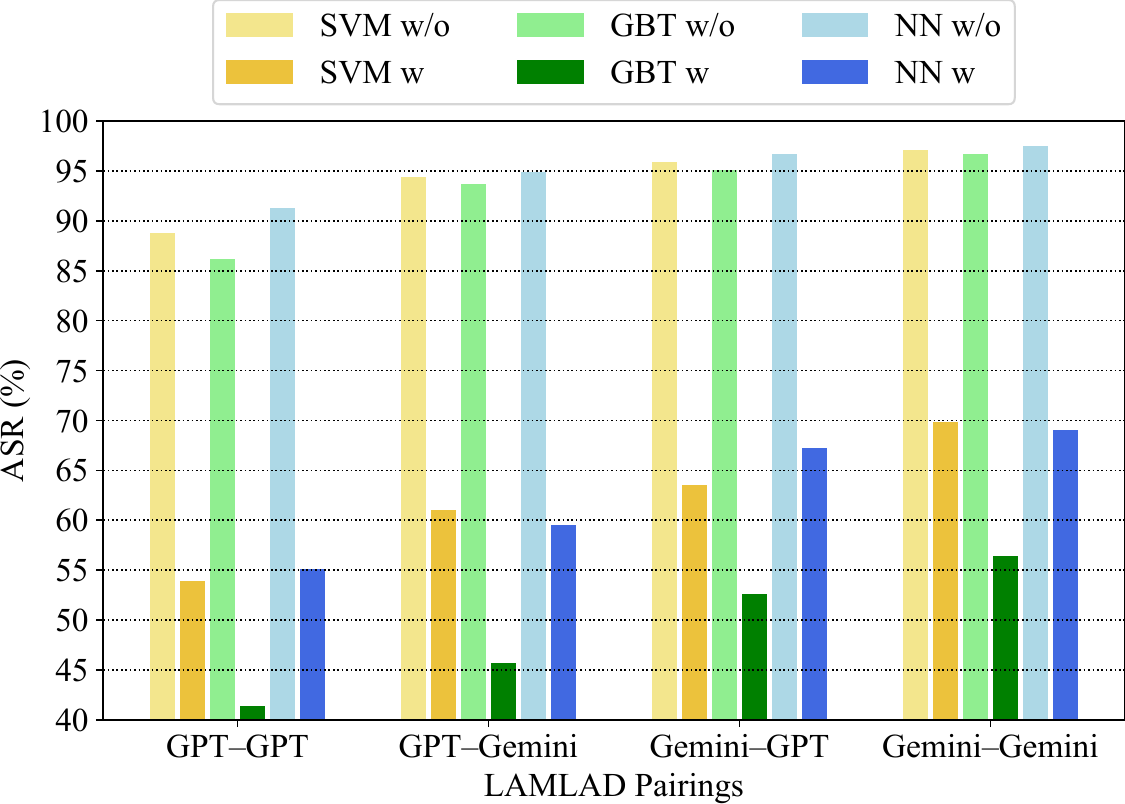}
    \caption{Adversarial training defense against LAMLAD on three ML detectors}
    \label{fig:adv_training}
\end{figure}

We apply adversarial training as a defense mechanism against LAMLAD on three ML-based malware detectors. Fig.~\ref{fig:adv_training} illustrates the reduction in ASR achieved by this defense across four LAMLAD pairings for the SVM, GBT, and NN models. In the figure, “w/o” denotes the LAMLAD attack without any defensive measures, whereas “w” indicates the results obtained after applying adversarial training.
Before the defense is applied, all ML detectors exhibit high vulnerability to LAMLAD, with ASR values reaching up to 97.52\% for the Gemini-Gemini configuration. Adversarial training consistently improves robustness across all evaluated models, reducing the ASR by more than 30\% in every case. Among the three detectors, GBT achieves the most pronounced improvement, with ASR reductions ranging from 40.27\% to 47.99\%. The NN model follows, exhibiting an average ASR reduction of 32.86\%. The SVM shows the smallest yet still substantial average reduction of 31.42\%.
Notably, all ML models remain more susceptible to LAMLAD configurations in which Gemini serves as the manipulator, particularly in the Gemini-Gemini pairing, which consistently yields the highest ASR even after the application of adversarial training. Overall, these results demonstrate that adversarial training is an effective defense strategy against LAMLAD across all evaluated ML detectors, achieving an average ASR reduction exceeding 30\%.

\section{Discussion}\label{sec:discussion}
This section discusses the limitations of LAMLAD and outlines potential directions for future research.

\subsection{Limitations}
All experimental results presented in this paper are obtained under an offline setting, in which LAMLAD generates adversarial examples prior to interacting with ML-based malware detectors. Consequently, the time overhead associated with adversarial example generation is not a critical factor in our current evaluation. In contrast, real-world and real-time deployment scenarios would impose stricter efficiency constraints, as the average generation latency is influenced by both the selected LLM and the integration of the RAG mechanism. Although RAG significantly improves response accuracy and factual consistency, it also incurs additional latency due to the retrieval process. Addressing this trade-off between accuracy and efficiency remains an important challenge and a promising direction for future work to enable practical real-time deployment of LAMLAD.

Moreover, LAMLAD in its current form targets ML-based malware detectors by manipulating Drebin features. While Drebin represents one of the most widely adopted static feature sets in both academic research and industrial applications, the generality of LAMLAD could be further enhanced by extending it to alternative static feature representations, such as MaMaDroid~\cite{onwuzurike2019mamadroid}, as well as to dynamic feature-based detection approaches.

\subsection{Future Work}
Future research will focus on both expanding the applicability of LAMLAD and developing more effective defense mechanisms against it. First, we plan to incorporate additional static and dynamic feature representations, enabling LAMLAD to adapt to a broader range of Android malware detection models. Second, we aim to explore more efficient RAG strategies and LLM fine-tuning techniques that improve response quality while reducing average generation latency, particularly in real-time attack scenarios. Finally, we intend to design novel defense strategies specifically tailored to LAMLAD, with the objective of strengthening the robustness of ML-based malware detectors against LLM-driven adversarial attacks. For example, analyzing the perturbation patterns and manipulation strategies employed by LLMs may help identify features that are frequently and easily modified without disrupting malicious functionality, thereby providing valuable insights for the development of more resilient defense mechanisms.

\section{Related Work}\label{sec:related}
ML-based Android malware detection methods can be broadly categorized into text-based, image-based, and graph-based approaches.
Text-based detection techniques rely on static or dynamic features extracted from APKs and represented in textual form. Drebin~\cite{ArpSHGR14} leverages a diverse set of static features, including permissions, API calls, and network addresses, to perform malware detection. Zhu et al.~\cite{multihead2023} further enhance static feature analysis by employing multi-head residual blocks to improve detection performance.

Image-based approaches transform APKs into visual representations, enabling the use of Convolutional Neural Networks (CNNs) for malware detection. Darwaish et al.~\cite{GCpaper} convert static features and application binaries into RGB images, allowing CNNs to learn discriminative patterns between benign and malicious applications. Similarly, the R2-D2 framework~\cite{hsien2018r2} introduces a CNN-based model that processes images generated from application bytecode, facilitating automated feature extraction without the need for manual feature engineering.

Graph-based methods model application behaviors using graph representations. MaMaDroid~\cite{onwuzurike2019mamadroid} constructs Markov chains from sequences of abstracted API calls, effectively capturing behavioral semantics of Android applications. Liu et al.~\cite{Enhancing2023} focus on identifying fine-grained malicious components within APKs by analyzing both structural and behavioral characteristics, thereby improving the granularity and precision of malware detection.

More recently, researchers have explored incorporating LLMs into Android malware detection frameworks. AppPoet~\cite{AppPoet2025} proposes a multi-view prompt engineering strategy in which static features of an APK are provided to an LLM to generate comprehensive behavioral summaries for malware classification. Qian et al.~\cite{LAMD2025} adopt a context-driven methodology that extracts security-critical code regions and applies tier-wise code reasoning with LLMs to progressively analyze application behavior.

Adversarial attacks against ML-based Android malware detection systems have also attracted considerable attention, revealing fundamental vulnerabilities in existing detectors. Grosse et al.~\cite{grosse2017adversarial} demonstrate that ML classifiers are susceptible to adversarial examples generated through minimal perturbations. Building upon this observation, HIV~\cite{HIV2020} injects carefully selected code snippets into benign regions of an APK’s bytecode and repackages malware, effectively evading detectors that rely on both syntactic and semantic features.
EvadeDroid~\cite{EvadeDroid2024} presents a practical black-box evasion attack by exploiting n-gram opcode similarity to identify benign APKs that are structurally similar to the target malware, from which functional code segments are extracted and incrementally injected. Lan et al.~\cite{AndroVenom2025} further introduce label spoofing attacks that poison crowd-sourced datasets by embedding subtle and difficult-to-detect malicious patterns into benign applications.

As a result of these evolving threats, defending against adversarial attacks has become a critical research focus. Adversarial training~\cite{goodfellow2014explaining} is a widely used defense strategy that enhances model robustness by augmenting training datasets with adversarial examples. SecureDroid~\cite{chen2017securedroid} combines feature selection with ensemble learning to strengthen ML-based Android malware detectors, proposing a feature selection scheme that accounts for both feature importance and manipulation cost. For image-based detectors, Lan et al.~\cite{Defensive2023ICC} propose defensive randomization, which applies randomized transformations to input images at inference time to mitigate the effectiveness of adversarial perturbations.

\section{Conclusion}\label{sec:conclusion}
In this paper, we introduced LAMLAD, a novel adversarial attack framework that exploits the generative and reasoning capabilities of LLMs to evade ML-based Android malware detection systems. LAMLAD adopts a dual-agent architecture comprising an LLM manipulator, which generates realistic feature-level perturbations while preserving the core malicious functionality, and an LLM analyzer, which steers the modification process toward successful evasion. To further enhance efficiency and contextual understanding, LAMLAD incorporates RAG into the LLM workflow. By operating on Drebin feature representations, LAMLAD enables stealthy and high-confidence attacks against widely deployed Android malware classifiers.
We evaluated LAMLAD on three representative ML-based Android malware detectors and benchmarked its performance against two state-of-the-art adversarial attack methods. In addition, we examined the effectiveness of different LLM configurations within the LAMLAD framework. Experimental results demonstrate that LAMLAD achieves an ASR of up to 97\%, requiring an average of only three attempts per adversarial example, underscoring its effectiveness, efficiency, and adaptability in realistic adversarial settings.
Moreover, we proposed an adversarial training-based defense strategy that reduces the ASR by more than 30\% on average, illustrating its effectiveness in strengthening the robustness of ML-based malware detectors against LAMLAD-driven attacks.

\bibliographystyle{ieeetr}
\bibliography{arxiv_LAMLAD/Ref}



\end{document}